\documentclass[12pt,a4paper]{article}
\usepackage[T2A]{fontenc}
\usepackage[utf8]{inputenc}
\usepackage[english]{babel}
\usepackage{amssymb}
\usepackage{amsmath}
\usepackage{graphicx}
\usepackage[small,sc,center]{titlesec}
\usepackage{indentfirst}
\usepackage[labelsep=period,labelfont=bf,small]{caption}    %
\usepackage{array, longtable, afterpage}
\usepackage{changes}
\usepackage{floatrow} 
\usepackage{rotating} 
\usepackage[small,sc,center]{titlesec}
\usepackage{xcolor}
\usepackage{authblk}
\usepackage{enumitem} 

\def\gtrsim{\ {\raise-.5ex\hbox{$\buildrel>\over\sim$}}\ }
\newcommand{\ms}{$M_\odot$}
\newcommand{\ls}{\mbox {$L_\odot$}}
\newcommand{\rs}{\mbox {$R_\odot$}}
\newcommand{\kms}{\:\mbox{km/s}}
\newcommand{\te}{\mbox{$T_\mathrm{eff}$}}

\renewcommand\baselinestretch{1.35}

\author[1]{Yu.A.~Fadeyev\thanks{eta.aql@gmail.com}}
\author[2]{A.G.~Kuranov}
\author[3]{L.R. Yungelson}
\affil[1,3]{Institute of Astronomy, Russian Academy of Sciences}
\affil[2]{Sternberg Astronomical Institute\\ M.V.~Lomonosov Moscow State University}

\title{\bf Elusive helium stars in the gap between subdwarfs and Wolf-Rayet stars~II.\\ Nonlinear pulsations of stripped helium stars}
\date{\small \today}
\begin{document}
    \maketitle
\textbf{Abstract} --- It is shown for the first time that the stripped helium stars with masses (2 -- 7)\,\ms\ which are formed
in close binary systems in the so-called case B of mass-exchange and retained low-mass hydrogen-helium
envelopes, experience  nonlinear radial pulsations.
Pulsations are excited by the $\kappa$-mechanism due to helium ionization.
The region of pulsational  instability extends over Hertzsprung-Russel diagram from the red 
giants branch to the region of effective temperatures $4.5\lesssim \log T_\mathrm{eff}\lesssim 4.7$.
Variations of stellar luminosity should be observed mostly in the ultraviolet. 
The amplitudes of  pulsations of the studied models reach $\Delta({\rm M_b})$=0.8 and increase, 
as the stellar radii $R$ decrease. 
Pulsation periods of stars with $\log T_\mathrm{eff} > 4$ range from 0.17 to 8.5 day and decrease with decreasing $R$.
The stars have substantially larger \te\ than their companions, which could be Be-stars.
They are components of relatively wide binaries with orbital periods up to several years.
The number of pulsating moderate-mass stripped helium stars in the Galaxy is $\simeq$\,1000.


\newpage
\section{introduction}
\label{sec:intro}
In the pioneering studies of the evolution of close binary stars (CBS), Kippenhahn et al. (1967a,b),
Paczy{\'n}ski (1967), Zi{\'o}{\l}kowski (1970), Giannone and Giannuzzi (1972)
found that at solar metallicity components of CBS with masses $\gtrapprox 2.8$\ms\ that
overflow Roche lobes at the stage of hydrogen shell-burning (so-called Case B of mass-exchange),
after cessation of 
the mass loss contract and transform into hot helium stars with thin hydrogen envelopes 
($\Delta(M_{\rm H})\lessapprox 1M_\odot$). 
These ``stripped helium stars''\footnote{The term is not quite correct, because the stars retain a 
fraction  of their hydrogen envelopes. 
Thus, they are ``half-stripped''.} spend in the core helium burning stage $\sim$10\% of the lifetime of
their progenitors at the main-sequence $t_{\rm MS}$.
Full-scale core-helium burning stage is preceded by a much shorter 
($\sim 0.01t_{\rm MS}$) stage of hydrogen-shell burning, in the course of which stellar radii decrease 
from tens and hundreds of \rs\ to (0.1 -- 1)\rs, and the luminosity drops by a factor of several.
Further studies have shown that stripped helium stars occupy a region of the Hertzsprung-Russell diagram
with $\log(T_\mathrm{eff})\approx$ (4.5 -- 5.0), $\log(L/\ls)\gtrsim$2.5.
It should be noted that solar metallicity stars with masses $\gtrsim 40$\,\ms\ may lose
much of their  hydrogen envelopes via stellar winds (Conti 1978).

For the helium stars in the contraction stage after completion of mass loss, with 
$\log(\te) \geq \log(\te)_{\rm ZAMS}+0.1$dex, Dutta and Klencki (2024) 
suggested a nickname ``puffed-up stripped stars''. 
We will use this term below.

Iben and Tutukov (1985, 1987) identified stripped helium stars with masses $\lessapprox 2$\,\ms\ with 
helium subdwarfs (sdB and sdO), while  Paczy{\'n}sky (1967) suggested that 
objects more massive than $\approx 7$\,\ms\ are Wolf-Rayet stars.
Note that at the time of writing there were known only about 20 binary subdwarfs (or candidates) with 
confirmed masses between 1\,\ms\ and 2\,\ms\ and known orbital periods (Wang et al. 2023;
Clement et al. 2024).
Typically, helium subdwarfs have masses $\lesssim$0.6\,\ms\ (Heber 2024).
The companions of massive subdwarfs in CBS are, as a rule, Be-stars
(see, e.g., Wang et al. 2023, Table ~9).
This suggests a prior exchange of matter, since the accretion of the matter pocessing  
angular momentum leads to a significant acceleration of rotation of companions 
of nascent stripped helium stars.

While binary  subdwarfs and Wolf-Rayet stars have been observed in the Galaxy,
helium stars with masses between 2\,\ms\ and 7\,\ms, even very rare,
were not known until very recently.
Meanwhile, helium star candidates -- components of detached CBS -- have been found
in the Large and Small Magellanic Clouds
(Villaseñor et al. 2023; Drout et al. 2023; G{\"o}tberg et al. 2023; Ramachandran et al. 2023, 2024).
This, despite the difference between metallicity of the LMC and SMC and the metallicity of 
the Galactic disk, allows us to claim that the theory of stellar evolution is correct and 
hot helium stars should also exist in the Milky Way.
The detection of hot helium stars is hampered by their low number 
(Yungelson et al. 2024; Hovis-Afflerbach et al. 2024) and 
the significant difference in the spectral characteristics of their host CBS.
The issue of detection of  helium stars was addressed in detail by  G{\"o}tberg  et al. (2018).

The only candidate stripped helium star in the Galaxy so far is HD~96670, which was previously 
considered as a  possible O/B star paired with a black hole.
However, Naz{\'e} and Rauw (2025) showed that the eclipses observed in the system
rule out the  black hole. 
Based on spectroscopic and photometric  observations, they suggested  that the
system harbours an O8.5 giant  with an $M\sim 4.5$\,\ms, $R\sim4.5$\,\rs,
 and $\log(\te)\sim$\,4.7 companion.
Such a star could be a remnant of a primary component of a CBS with an
initial mass close to (15 -- 16)\ms\ (Yungelson et al. 2024).
Irrgang, Przybilla and Meynet (2022) suggest that in the $\gamma$\,Col system
the  bright B-class component  with mass $\simeq$4\ms,
$T_{\mathrm eff} = 15\,570 \pm 320$\,K and $\log(g)=3.3\pm 0.1$
is an inflated stripped star.
To the possible loss of mass by this star and its large radius,
in addition to the low $T_\mathrm{eff}$ and $\log(g)$, point anomalously high
nitrogen abundance at the surface.

Moreover, several detached CBS harbouring subdwarfs with large radii and 
Be-stars -- LB-1 (Shenar et al. 2020a; Lennon et al. 2021; El-Badri and  Quataert 2021),
HR~6819 (Bodensteiner et al. 2020),
as well as low-mass semi-detached systems HD~15125 (El-Badri et al. 2022)
and V315~Cas (Zak et al. 2023) with parameters of donors and accretors 
similar to LB-1 and HR~6819 have been found in the Galaxy.  
A similar detached system NGC~1850~BH1 has been found in LMC (El-Badri and Burdge 2022).
Relatively low $T_\mathrm{eff}$ of the low-mass components in these CBS, their
positions in the Hertzsprung-Russell and $\log(T_{eff}) - \log(g)$ diagrams  
suggest that they are either inflated stripped stars or stars with helium-depleted cores  
expanding in the helium-shell burning stage (El-Badri and Burdge 2022). 
Stars like HD~15125 and V315~Cas may be their immediate progenitors (El-Badri et al. 2022).
Rivinius et al. (2025) confirmed that the components of LB-1 and HR~6819 are Be-stars and subdwarfs. 
They also suggested that in the four systems harbouring Be-stars and subdwarfs which they 
observed, the subdwarfs have only recently completed mass loss and still have extended radii.

Detection of intermediate mass stripped helium stars and/or their progenitors,
the absence of which among the observed stars
has been a ``mystery'' for about fifty years, provides a significant opportunity
to test the theory of stellar evolution and to study stellar pulsations.
These stars are of high interest, since if they have masses exceeding $\simeq 2.2$\,\ms\
they could be the precursors of Type~Ib and Type~Ic supernovae (Habets 
1986; Woosley, Langer, and Weaver 1995). At the lower mass end they, possibly, may be 
precursors of electron-capture and peculiar type SN\,Ia supernovae (e.g., 
Chanlaridis et al. 2022).

A systematic study of the  hypothetical Galactic population of intermediate mass stripped 
helium stars (Z=0.02) was carried out by Yungelson et al. (2024).
They considered CBS with ZAMS masses of primary components 
from 4 to 25\,\ms,  mass ratios of components $q=M_2/M_1$ = 0.6, 0.8, 0.9, and the range of the
initial orbital periods $P_{\mathrm orb}$=(2 -- 1000)\,day.

Hovis-Afflerbach et al. (2024) performed a similar study for the stars with ZAMS
masses from 2 to 18.7\ms, $q=0.8$, and $P_{\mathrm orb}$ from 3 to 31.5 days,
with $Z=$0.014, 0.006, 0.002, and 0.0002.
In both papers, a hybrid population synthesis was performed, taking into account the results 
of detailed evolutionary calculations using the MESA code
(see Jermyn et al. (2023) and references therein).
For $Z=0.02$ and $Z=0.014$, the populations of objects with masses (2 -- 7)\,\ms\ were 
estimated, respectively, as $\simeq$3000 by Yungelson et al. (2024)
and $\simeq$\,4000 by Hovis-Afflerbach et al. (2024). This should be recognized as a 
reasonable agreement given the acceptable differences in the parameters of stellar models 
and population synthesis.
Previously, Shao and Li (2021), using the BSE population synthesis code (Hurley, Tout, 
and Pols 2002), estimated the population of stripped helium stars in the Galaxy as 
$\sim 10^3$.

This paper is a continuation of the study by Yungelson et al.  (2024; Paper I) on the 
modeling of stripped helium stars of moderate masses in the Milky Way.
We have attempted to investigate the nonlinear pulsations of the 
(2 -- 7)\,\ms\ (masses of progenitors 9 to 16 \ms) stars at the stage lasting from 
departure from the Roche lobe 
to the core helium exhaustion and to estimate the possible number of such objects.
For this purpose, we considered hydrodynamical models of pulsations of the remnants of 
donors in the CBS with different ZAMS orbital periods.

In Sec.~\ref{sec:stars} we describe the calculated evolutionary models and their structure.
In Sec.~\ref{sec:puls} the methodology of the pulsations calculations is described  and 
their results are summarized.
An estimate of the number of nonlinearly pulsating stars using population synthesis method
is presented in Sec.~\ref{sec:popsynt}.
The results are discussed in Sec.~\ref{sec:disc}.

\section{EVOLUTIONARY MODELS OF STRIPPED HELIUM STARS}
\label{sec:stars}
\begin{figure}[t!]
\includegraphics[width=0.8\textwidth]{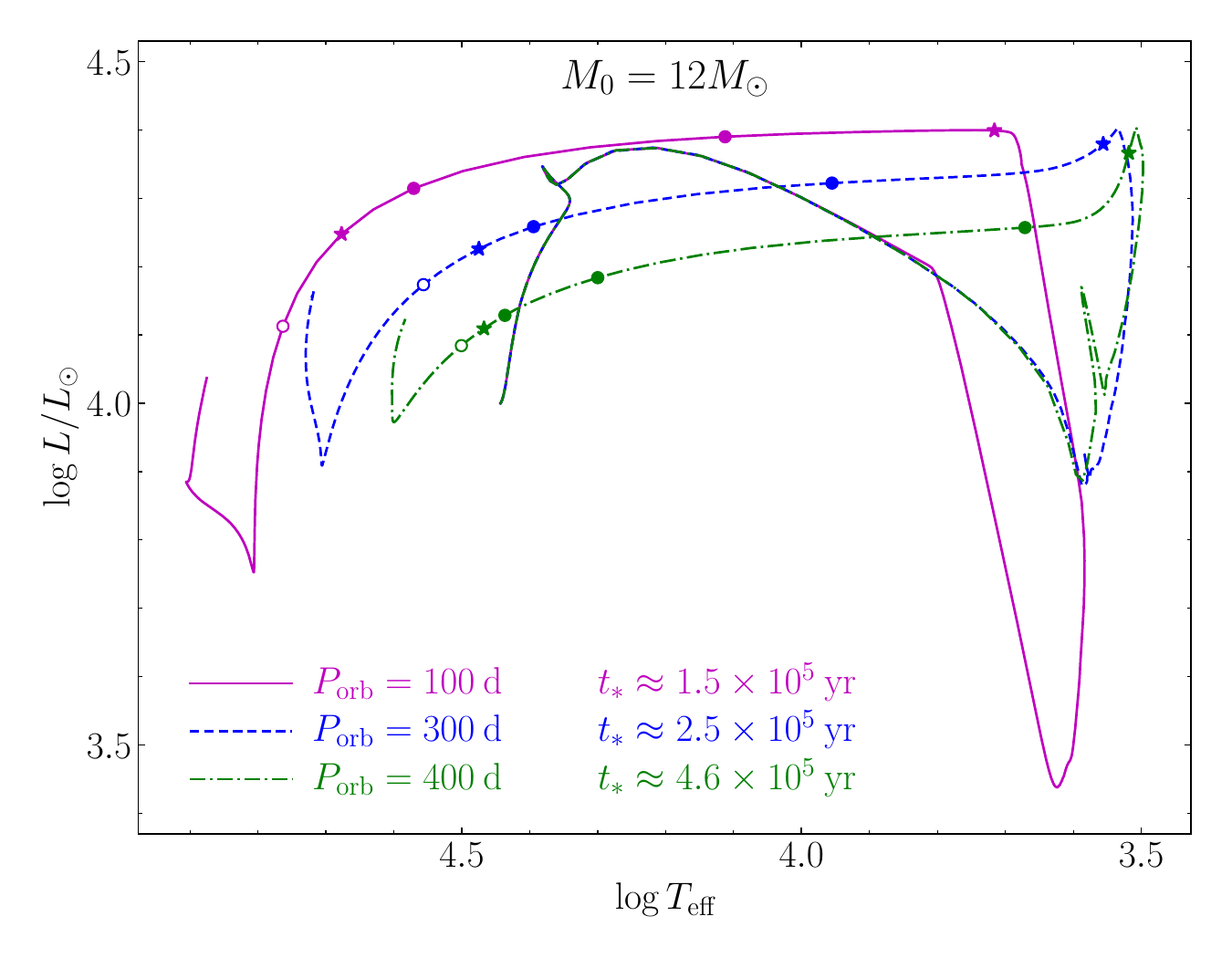}
\caption{Evolutionary tracks of the primary components of CBS with initial mass $12M_\odot$
in the systems with an initial orbital period of 100~day (solid line), 300~day
(dashed line), 400~day (dash-dotted line). Asterisks on the tracks indicate
the beginning of He burning in the core (in the cold region of the diagram) and the end of the pulsation stage (in the hot region).
 Filled and empty circles mark, respectively,
pulsating and stable models, respectively. $t_\star$ is duration of the pulsation stage.
}
\label{fig1}
\end{figure}
We have investigated the nonlinear pulsations of stars with ZAMS masses 
9, 12, and $16M_\odot$.
According to Yungelson et al. (2024), stripped helium stars mass range $(2 - 7)M_\odot$ 
corresponds to stars with initial masses up to $\approx 20M_\odot$ (depending on 
$P_\mathrm{orb}$ at the instant of Roche lobe overflow by the donor), but it is clear that the 
population of stars with initial masses exceeding  $16M_\odot$ is negligible 
compared to the less massive stars.
Assumptions concerning convective boundary mixing, stellar winds, and population synthesis
parameters are described in Paper I. 
The only significant difference is that for the system with the 
primary component mass $M_{1,0}$=16\,\ms\ the efficiency of convective overshooting at the 
boundary  of the stellar hydrogen core was not constrained, but a fixed parameter
of exponential overshooting $f_{ov}=0.004$ was applied. This leads to a slightly larger mass of the stripped helium star.
For CBS with initial masses of primary components 9 and 12\,\ms\, the initial 
value of the mass ratio of component $q$ was taken as 0.8, while for the 
system with $M_{1,0}$=16\,\ms\ it was equal to 0.9.

Following Packet (1981), we assumed that mass exchange occurred conservatively until accretor's
equatorial angular velocity of rotation attained 95\% of the critical one 
($\omega_\mathrm{cr}$). 
Then, accretion was limited by $0.95\times \omega_\mathrm{cr}$, while the 
excess of the matter was expelled from the system by the wind enhanced by rotation, taking away 
specific angular momentum of the accretor.
Angular velocity of rotation attains its limit after the transfer of  only 5\% to 10\% of the 
donor mass, so the mass of the accretor remains essentially unchanged.
But, importantly, its equatorial rotation velocity reaches hundreds of km/s and it must 
become an Oe/Be-star.

Evolutionary and pulsational characteristics of the models are summarized in 
Table~\ref{t:models}.
Due to the considerable amount of computer time required by calculation of the hydrodynamical models, we 
have limited ourselves by a detailed consideration of pulsations of the primary components of CBS with the ZAMS mass  
$M_0=12M_\odot$ and initial periods $P_0$=100, 300, 400 day.
For the models with $M_0=9$ and $16\,M_\odot$ we considered models with $P_0$=100 and 200 day, 
respectively.
Table 1 also lists the masses, luminosities, and effective temperatures of the companions
to the modeled stars.
Note that the latter characteristics almost do not change  during the phase in which 
stripped helium stars experience pulsations.
In Fig.~\ref{fig1} we show the tracks of stars with $M_0=12M_\odot$ and mark on them positions of models for which 
hydrodynamic  computations of stellar pulsations were performed.

\afterpage{\clearpage\begin{sidewaystable}
	\centering
	\caption{Evolutionary and pulsational characteristics of hydrodynamical models.}
	\vskip -4pt
	\small
\label{t:models}
	\begin{tabular}{c|c|c|c|c|c|c|c|c|c|c|c|c|c|c|c}

	\hline\hline
		$M_0$,   & $P_0$, & $t_{\mathrm{ev}}$, & $M_1$, & $P_\mathrm{orb}$,& $X_s$  & $Y_s$    & $\log(L/L_\odot) $ & $\log(T_{\mathrm{eff},1})$ & $\Pi,$   & $\eta$ & $\Delta U_s$&
		$\Delta M_\mathrm{bol},$  & $M_2$,  & $\log(L_2/L_\odot) $ & $\log(T_\mathrm{eff,2})$\\[-8pt]
		$M_\odot$       & day   &  $10^6$ yr      &$M_\odot$   &day &        &          &                      &               &    day            &        &         & st. mag.
		&    $M_\odot$         &        & \\
		\hline
		9     & 200 &  26.748  &  1.995 & 1782 &  0.478 &   0.502 &  3.77 &   4.017 &  2.899    &  -0.333 &      &       & 7.32 & 3.46 & 4.27 \\ [-0.2em]
		      &     &  26.909  &  1.992 & 1783 &   -"-  &   -"-   &  3.73 &   4.222 &   0.747   &   0.076 &  162 &  0.44 &      &      &      \\[-0.2em] 
		      &     &  27.061  &  1.989 & 1783 &   -"-  &   -"-   &  3.68 &   4.326 &   0.299   &   0.041 &  218 &  0.58 &      &      &      \\[-0.2em] 
		      &     &  27.208  &  1.986 & 1786 &    -"- &    -"-  &  3.64 &   4.389 &   0.168   &  -0.014 &      &       &      &      &      \\
		\hline
		12    & 100 &  15.435  &  2.633 & 807.6&  0.266 &   0.715 &  4.40 &   3.766 &  66.659   &  -1.403 &      &       & 9.73 & 3.88 & 4.33 \\ [-0.2em]
		      &     &  15.454  &  2.629 & 808.2&  0.265 &   0.716 &  4.39 &   4.113 &   6.067   &   0.083 &  133 &  0.23 &      &      &      \\ [-0.2em]
		      &     &  15.538  &  2.600 & 812.1&  0.255 &   0.726 &  4.31 &   4.571 &   0.142   &   0.001 &  265 &  0.57 &      &      &      \\ [-0.2em]
		      &     &  15.642  &  2.575 & 815.5&  0.247 &   0.733 &  4.11 &   4.763 &   0.024   &  -0.015 &      &       &      &      &      \\ [-0.2em]
		      & 300 &  15.630  &  2.811 & 2456 &  0.422 &   0.558 &  4.32 &   3.955 &  13.155   &   0.087 &   47 &  0.21 &  9.69 & 3.88 & 4.34\\ [-0.2em]
           &\textbullet&  15.709  &  2.794 & 2463 &   -"-  &  -"-    &  4.26 &   4.394 &   0.559   &   0.561 &  272 &  0.35 &      &      &      \\ 
		      &     &  15.799  &  2.788 & 2466 &  -"-   &  -"-    &  4.17 &   4.556 &   0.108   &  -0.083 &      &       &      &      &      \\[-0.2em] 
		      & 400 &  15.783  &  2.940 & 2947 &  0.497 &   0.484 &  4.24 &   3.974 &   8.507   &   0.079 &   65 &  0.16 & 9.68 & 3.88 & 4.34 \\ [-0.2em]
		      &     &  15.867  &  2.930 & 2952 &  -"-   &  -"-    &  4.18 &   4.300 &   0.910   &   0.446 &  256 &  0.30 &      &      &      \\ [-0.2em]
		      &     &  15.953  &  2.919 & 2957 &  -"-   &  -"-    &  4.13 &   4.436 &   0.251   &   0.061 &  262 &  0.60 &      &      &      \\ [-0.2em]
		      &     &  16.036  &  2.918 & 2958 &  -"-   &  -"-    &  4.08 &   4.501 &   0.134   &  -0.015 &      &       &      &      &      \\ 
		\hline
		16    & 100 &  12.534  &  6.636 & 235.4 &  0.276 &  0.705 &  5.09 &   4.579 &   8.540   &   0.557 &  258 &  0.60 & 14.27 & 4.62 & 4.36 \\[-0.2em] 
		      &     &  12.542  &  6.615 & 235.9 &  0.272 &  0.709 &  5.08 &   4.645 &   3.893   &   0.177 &  300 &  0.65 &      &      &      \\[-0.2em] 
		      &[    &  12.549  &  6.595 & 236.4 &  0.268 &  0.713 &  5.06 &   4.701 &   1.660   &   0.131 &  581 &  0.81 &      &      &      \\-0.2em]  
		      &     &  12.556  &  6.576 & 236.8 &  0.264 &  0.717 &  5.05 &   4.747 &   0.777   &   0.000 &  472 &  0.71 &      &      &      \\ 
		\hline
	\end{tabular}

\vskip  2 mm
\begin{minipage}{\textwidth}
\renewcommand\baselinestretch{0.7}
\begin{flushleft}
{\small Notes. $M_0$, $P_0$ -- ZAMS mass and orbital period of the system, respectively,
$t_{\mathrm{ev}}$ -- the age of the model,
$M_1$ -- mass of the model,
$P$ -- orbital period of CBS,
$X_s$, $Y_s$ -- surface mass abundances of hydrogen and helium,
$\log(L/L_\odot)$ -- luminosity of the model,
$\log T_\mathrm{eff}$ -- effective temperature of the model,
$\Pi$  -- period of pulsations,
$\eta = \Pi d\ln E_\mathrm{K}/dt$ -- the rate of increase ($\eta > 0$) or decrease ($\eta < 0$)
of kinetic energy of pulsations $E_\mathrm{K}$,
$\Delta U_s$ -- the amplitude of variation of gas velocity,
$\Delta M_\mathrm{bol}$ --  the amplitude of variations of the bolometric stellar luminosity at the outer boundary of the hydrodynamic model, 
$M_2$, $\log(L_2/L_\odot)$ and $\log(T_{\mathrm eff,2})$ -- mass, decimal logarithms of luminosity, and 
effective temperature of the companion of stripped helium star, respectively.
} 
\end{flushleft}
\end{minipage}
\end{sidewaystable}
\nopagebreak
\clearpage}

\section{RADIAL PULSATIONS OF STRIPPED HELIUM STARS}
\label{sec:puls}.
\begin{figure}[t!].
\includegraphics[width=0.8\textwidth]{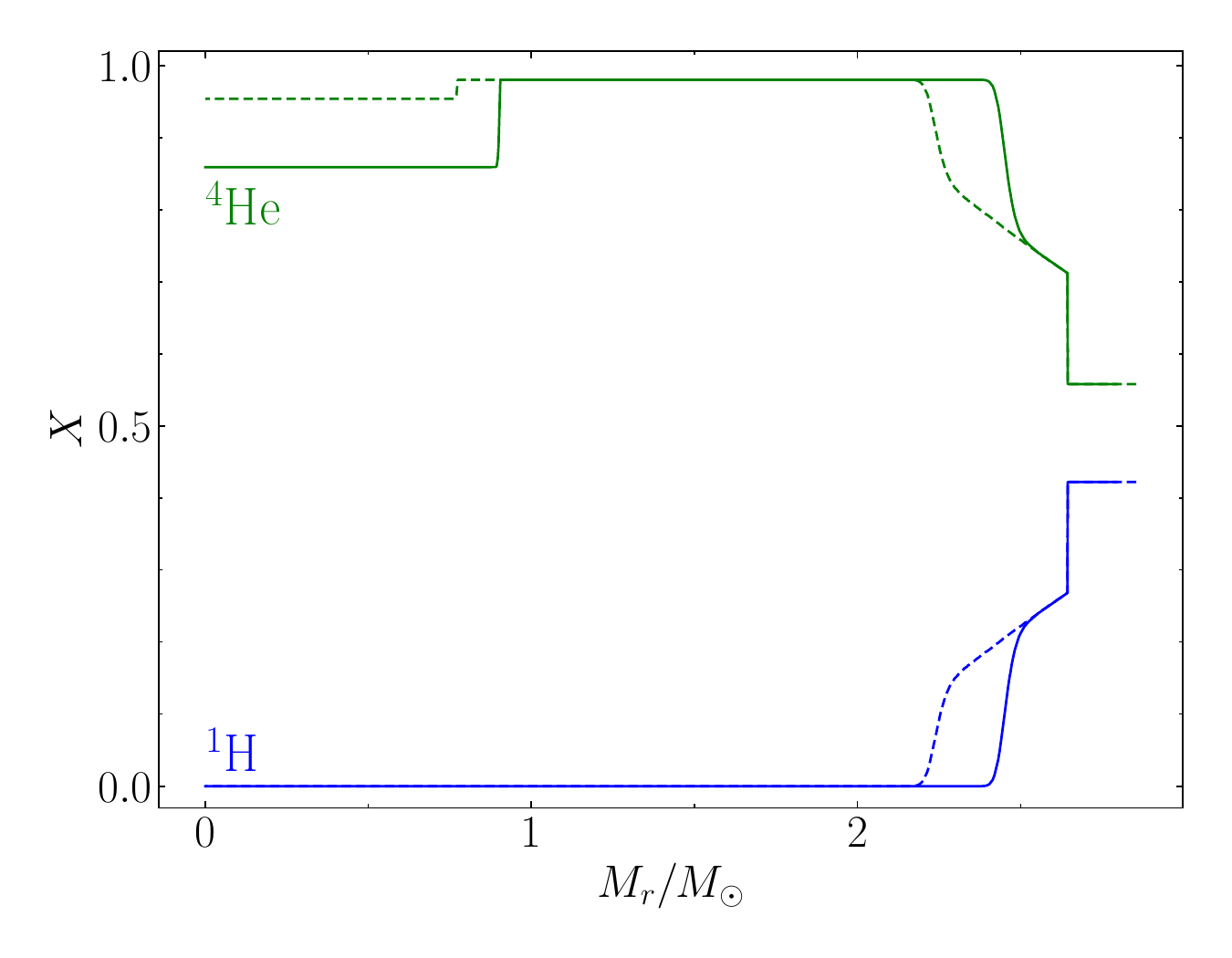}
\caption{Distribution of helium and hydrogen abundances in the puffed-up star
at the beginning of helium burning in the core (dashed lines) and at the pulsation instability boundary (solid line).}
\label{fig2}
\end{figure}

By the epoch when the loss of matter is complete, helium burning starts in the cores of the 
progenitors of stripped helium stars. 
At the stage when the stars are pulsing, helium abundance in the cores  $Y_c$ decreases further by 
$\simeq 0.1$.
However, throughout this stage hydrogen-shell burning remains the dominant energy source. 

Due to the loss of a part of stellar envelope, the matter affected by the
earlier nucleosynthesis and, therefore, characterized by a deficit of hydrogen and an excess of helium
appears in the outer layers of the star.
As an illustration, Fig.~\ref{fig2} shows the profiles of hydrogen and helium abundances in two
models of a puffed-up stripped  helium star -- the remnant of a star with ZAMS mass $12M_\odot$  and 
initial orbital period of 300~day.
The first of these models (dashed lines in Fig.~\ref{fig2}) is in the early stage of core helium burning.
Effective temperature of the star is $T_\mathrm{eff}=3.6\times 10^3\:\textrm{K}$ and in the HR-diagram 
it is located close to the  luminosity maximum of the track. In Fig.~\ref{fig1} this model is marked by an asterisk.
The effective temperature of another model (solid lines in Fig.~\ref{fig2})
is $T_\mathrm{eff}=2.5\times 10^4\:\textrm{K}$.
In  Table ~\ref{t:models}, this model is labeled by a bullet (\textbullet).
Among  hydrodynamic models in this evolutionary sequence, this model is the last one,
which is unstable with respect to radial pulsations.

It is known that with increase of the helium abundance, the boundary of the region in the HR diagram 
populated by the radially pulsating stars shifts toward high effective 
temperatures, which  significantly exceed $10^4$\,K (Fadeyev and Novikova 2003).
The stage of evolution when the stars pulsate lasts for hundreds of thousands of years (see 
$t_\star$ in Fig.~\ref{fig1} for approximate estimates). Thus it is of great interest to consider the
origin of pulsations and to determine the periods and amplitudes of light variations.

For the evolutionary models listed in Table~1 we have carried out hydrodynamic computations of nonlinear stellar oscillations.
The system of the equations of radiation hydrodynamics and the time--dependent convection as well as parameters of
the convection theory (Kuhfu\ss\ 1986) are discussed in the paper by Fadeyev (2013).
In a contrast to the evolutionary calculations based on the adaptive difference scheme, the equations of hydrodynamics
were solved on the fixed Lagrangian grid.
Initial values of the grid functions of hydrodynamic models were computed by cubic spline interpolation of the
corresponding variables of evolutionary models.
Thus, the hydrodynamic model of stellar pulsations is consistent with the model of the evolutionary sequence
due to the same spatial distributions of physical variables and isotope abundances.
The inner boundary of the hydrodynamic model was set in the layers with temperature
$10^7\,\textrm{K}\lesssim T\lesssim 2.5\times 10^7\,\textrm{K}$ \footnote{Progenitors of stripped helium stars are the
stars with mass $\gtrsim 12M_\odot$ and hydrogen burning in the CNO--cycle takes place at the temperature
$T\sim 4\times 10^7\,\textrm{K}$.}.
The equations of hydrodynamics do not take into account thermonuclear energy sources
so that one of the inner boundary condition implies $L_0 = L$, where $L_0$ is the luminosity at the
inner boundary of the hydrodynamic model and $L$ is the luminosity of the evolutionary model.

Solution of the Cauchy problem for the equations of hydrodynamics describes the self--excited stellar oscillations
where interpolation errors arising in the initial conditions calculations act as 
initial hydrodynamic perturbations.
In such a way, integration of the equations of hydrodynamics leads to the solutions of two types.
The first type corresponds to decaying oscillations when the star is stable against radial oscillations,
whereas the second one describes oscillations with exponentially increasing oscillation amplitude.
Pulsations of helium stars are driven by the $\kappa$--mechanism in the helium ionization zones.
The amplitude growth ceases with subsequent transition to the limiting amplitude (saturation of the $\kappa$--mechanism)
when the compressed gas approaches full helium ionization, so that stronger compression is accompanied by
decrease of the opacity (Christy 1966; Cox et al. 1966).
The pulsation period of large--amplitude oscillations varies with time within $\approx 10$ per cent,
but the limiting amplitude evaluated for the large number of cycles remains almost constant.

\begin{figure}[t!]
\centering
\includegraphics[width=0.8\textwidth]{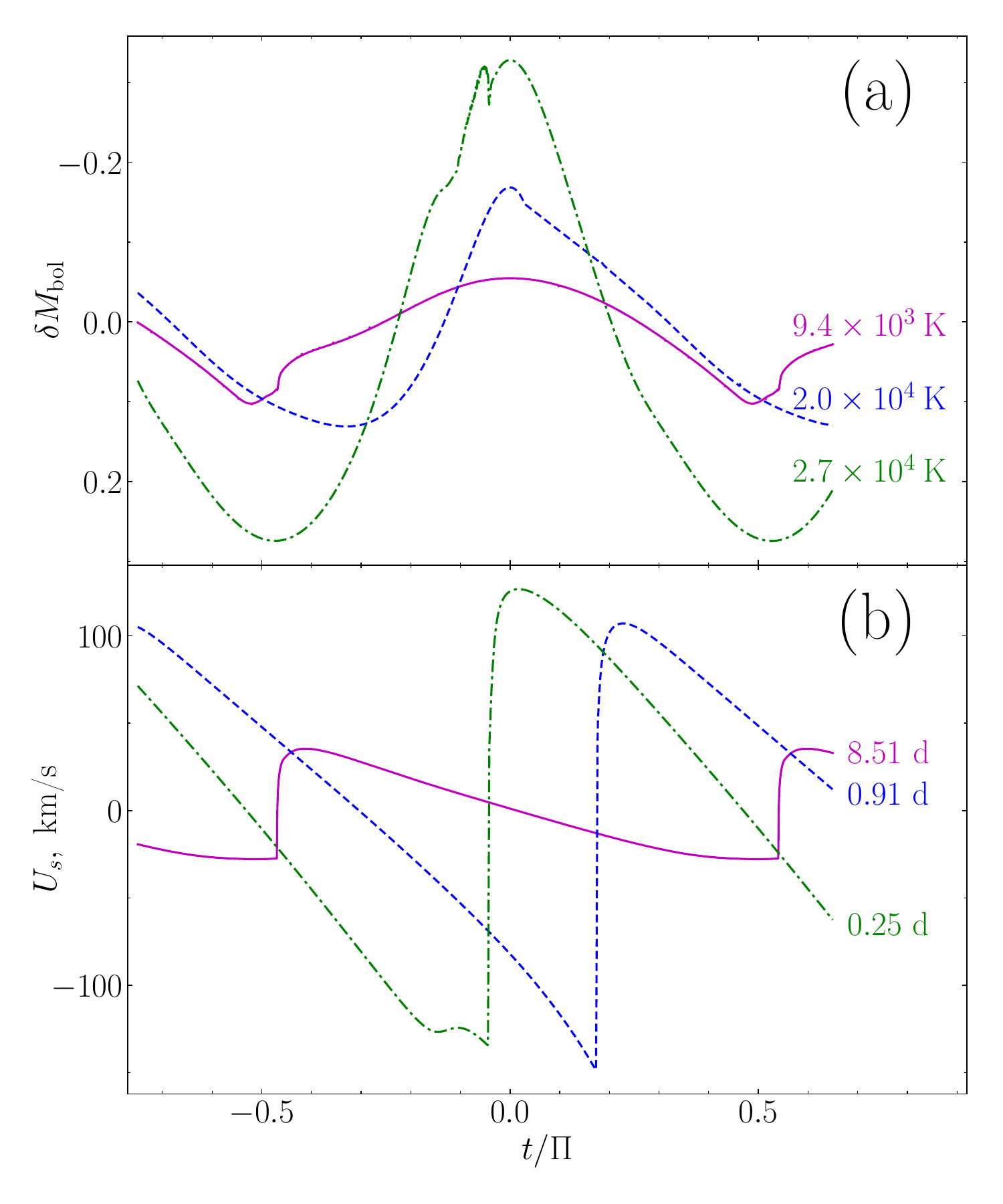}
\caption{Variations of the bolometric magnitude (a) and the gas flow velocity at the outer boundary (b)
in three hydrodynamic models of the evolutionary sequence $M_\mathrm{ZAMS}=12M_\odot$, $P_\mathrm{orb}=400$~d.
The plots are marked by values of the equilibrium effective temperature (a) and the pulsation period (b).  }
\label{fig3}
\end{figure}

The pulsation period of the hydrodynamic model $\Pi$ was calculated using the discrete Fourier transform of the
kinetic energy of pulsation motions
\begin{equation*}
 E_\mathrm{K}(t) = \frac{1}{2} \sum_{j=1}^N U_j(t) \Delta M_j,
\end{equation*}
where $U_j$ and $\Delta M_j$ are the gas flow velocity and the mass of the Lagrangian interval of the $j$--th
zone, $N=400$ is the total number of Lagrangian intervals.
The kinetic energy $E_\mathrm{K}(t)$ was computed for a time covering a few hundred pulsation cycles,
so that the period $\Pi$ was evaluated with relative error less than one per cent.

The main properties of hydrodynamic models are presented in Table~1, where $\Pi$ is the pulsation period,
$\eta = \Pi d\ln E_\mathrm{K}/dt$ is the growth ($\eta > 0$) or decay ($\eta < 0$) rate of the kinetic
energy $E_\mathrm{K}$, $\Delta U_s$ and $\Delta M_\mathrm{bol}$ are the amplitudes of the gas flow velocity
and the bolometric magnitude at the outer boundary of the model.

Results of nonlinear pulsation calculations are illustrated in Fig.~3, where variations of the bolometric
magnitudes and the surface gas velocity are plotted for three hydrodynamic models of the evolutionary
sequence $M_\mathrm{ZAMS}=12M_\odot$, $P_\mathrm{orb}=400$~d.
For the sake of graphical convenience, the variations $\delta M_\mathrm{bol}$ are plotted in respect 
of the average (i.e. equilibrium) bolometric magnitude of the evolutionary model.

As seen in Fig.~3, notwithstanding the significant radial velocity amplitude at the stellar surface
($\Delta U_s\sim 200$~km/s), the amplitude of light variations does not exceed a half of magnitude.
This feature of pulsating helium stars is quite different from that, for example, in Cepheids
due to the absence of the hydrogen ionization zone in hot pulsating helium stars.
Increase of the light amplitude $\Delta M_\mathrm{bol}$ with increasing effective temperature is due to
displacement of helium ionization zones to the outer layers with larger pulsation amplitude.
It should be noted that increase of the effective temperature in helium stars is accompanied by
decreasing mass of helium ionization zones.
Increase of $T_\mathrm{eff}$ ultimately leads to a cessation of pulsations when the total mechanical work
$\oint PdV$ done over the cycle in the helium ionization zones becomes less than the total
mechanical work in deeper layers of fully ionized helium suppressing pulsations.
For models of the evolutionary sequence $M_\mathrm{ZAMS}=12M_\odot$, $P_\mathrm{orb}=400$~d presented in
Fig.~3 the edge of the pulsation instability region corresponds to the effective temperature
$T_\mathrm{eff}\approx 3\times 10^4$~K.
At this phase of stellar evolution the central temperature is $1.6\times 10^8$\,K and the energy generation rate
due to helium burning is $\sim 40\%$ of the energy released by the hydrogen burning shell source.
In other words, core helium burning gradually replaces hydrogen shell burning and becomes the primary
energy source.

\section{the number of pulsating stripped helium stars}
\label{sec:popsynt} 
\begin{figure}[t!]
\includegraphics[width=0.8\textwidth]{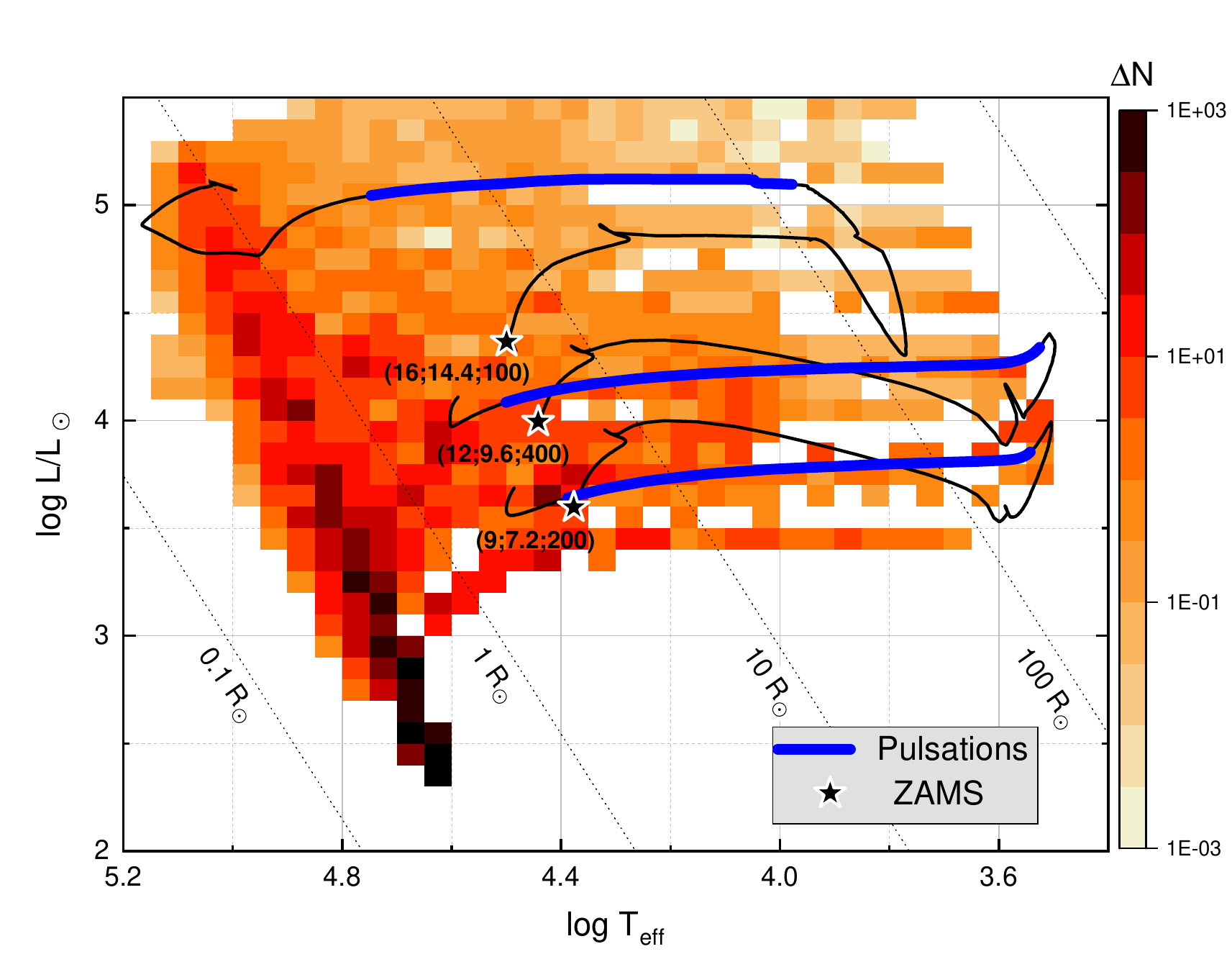}
\caption{Stripped helium stars in the HR-diagram (Yungelson et al. 2024).
Colour scale shows the number of stars per
$\Delta(\log(\te)) \times \Delta(log(L/\ls))=0.05 \times 0.1$ cell.
Overplotted by solid lines are examples of stellar evolutionary tracks for which
pulsations calculations were performed (see Table ~\ref{t:models}).
The numbers by the tracks indicate ZAMS masses of components and orbital periods. 
The parts of the tracks where pulsations are possible are highlighted in blue.
}
\label{fig4}
\end{figure}

Figure~\ref{fig4} shows distribution of the number of stripped helium stars with masses (1 -- 7)\,\ms\ 
over HR-diagram (Yungelson et al. 2024).
The region occupied by pulsating stars can be approximately bounded by red giants branch at the 
``cold'' side and the line of constant radii
$R\approx 4$\,\rs\ at the ``hot'' side,  i.e., it encompasses not only the ``puffed-up'' stripped 
helium stars with hydrogen-shell burning dominating energy release, 
but also stars in which core He-burning begins to play a significant role.
Current population of Galactic pulsating helium stars may be estimated as $\approx$1000, i.e.,
they comprise  about (25 -- 30)\% of all helium stars with masses from 2 to 7\,\ms\ (for star 
formation 
rate in the Galaxy  2\,\ms/yr (Chomiuk and Povich, 2011)). 

Note that the Galactic stripped helium stars with 
$\log(T_{\mathrm eff}) \gtrapprox 4.4$ and $\log(L/\ls) \gtrapprox 4.9$ are identified with 
nitrogen Wolf-Rayet stars (e.g., Shenar et al 2020b).

We considered above, like in Paper~I, only stars formed by stable mass-loss.  Unstable mass-loss 
should, typically, result in formation of common envelopes. 
The possibility of formation of He-stars as a result of evolution in common envelopes was not 
addressed, due to 
the absence of theoretical models of the structure of stars that ``survived'' in the common envelopes.
This circumstance limits the initial periods of the precursors of considered stripped helium stars 
in CBS by (200 -- 500)\,day,
depending on the initial mass and $q$. This, possibly underestimates the number of the latter.
Using ``standard'', but in fact arbitrary, values of so-called ``common envelope efficiency''
and binding energy parameter of the stellar envelopes
$\alpha=1$ and $\lambda$=0.5, respectively,
Hovis-Afflerbach et al. (2024) estimated that the fraction of stripped helium stars forming in the
common envelopes does not exceed $\simeq$20\%.

\section{discussion and conclusions}
\label{sec:disc}
In the present study, we found a new class of pulsating stars -- stripped helium stars with masses 
(2 -- 7)\,\ms, formed as a result of a stable mass-loss in CBS in the so-called case B of 
mass-exchange (RLOF by the  more massive component of the system in the hydrogen-shell burning stage). 
Stripped helium stars start to pulsate after cessation of mass transfer and continue until their
effective temperature becomes $T_\mathrm{eff}\gtrsim (2.5 - 5)\times 10^4\:\textrm{K}$.
The upper limit of the effective temperatures increases with increasing stellar mass.
The characteristic values of pulsation periods $\Pi$ range from a few hours to a few days.
The number of pulsating helium stars in the Galaxy is estimated to be $\simeq$1000.

Pulsations are excited by the $\kappa$-mechanism due to the opacity maximum associated with the helium 
ionisation.
Remarkably, in massive stars, $\kappa$-mechanism works in the region of red giants in the HR-diagram, 
while for stripped helium stars its region of action forms a kind of a ``band'' that encompasses both the region 
of red giants and the part of the HR-diagram that stretches to ZAMS and a  part of a 
hotter region, i.e., the zone in which the action of  $\kappa$-mechanism
is associated with Z--bump in layers with temperature from $2\times 10^5$~K to $3\times 10^5$~K
(e.g., in stars of the $\beta$~Cep type).

A distinctive feature of the pulsations is the large amplitude (up to several hundred \kms)
of the variation of the gas flow velocity at the outer boundary, while the amplitude of the bolometric 
luminosity change
$\Delta M_\mathrm{bol}$ does not exceed one stellar magnitude.
Together with the high effective temperature of the majority of pulsating helium stars
($10^4\:\textrm{K}\lesssim T_\mathrm{eff}\lesssim 5\times 10^4\:\textrm{K}$),
the above estimates of $\Pi$, $\Delta M_\mathrm{bol}$, and $\Delta U_s$
can be used as a criterion by which a detected variable star
can be classified as a stripped helium star with outer layers lost via RLOF.
Also, significant variations of the gas flow velocity in the outer layers
of a pulsating star suggest generation of periodic shock waves in the stellar
atmosphere.
Taking into account relatively high hydrogen abundance  in the outer layers of the 
considered models of
helium stars (see Table~1), one can expect that one of the indicators of nonlinear stellar pulsations will be
hydrogen emission lines appearing at the shock front during each oscillation cycle
close to the luminosity maximum.

It should be noted that pulsating stripped helium stars can be components of CBS with
orbital periods up to 7 -- 8 yr and that their companions may be Be-type stars.

\vskip 0.5cm
{\small The authors acknowledge K.A. Postnov for his attention to this study and important advice.
They appreciate critical remarks of the referees which allowed to correct inaccuracies and to
improve the presentation.}

\section*{funding}

This work was supported by ongoing institutional funding. No additional grants 
to carry out or direct this particular research were obtained.
A.K. work was performed according to the state assignment to the  M.V. 
Lomonosov Moscow State University. 

\section*{conflict of interests}
The authors declare that they have no conflicts of interests.

\section*{references}

{\small
\begin{enumerate}[itemsep=0em]

\item J.~Bodensteiner, T.~Shenar, L.~Mahy, M.~Fabry, P.~Marchant, M.~Abdul--Masih, G.~Banyard, D.M.~Bowman, K.~Dsilva, A.J.~Frost, C.~Hawcroft, M.~Reggiani, and H.~Sana,
      Astron. Astrophys. \textbf{641}, A43 (2020).

\item Chanlaridis, S., Antoniadis, J., Aguilera-Dena, D.~R. et al., Astron. Astrophys., 668, A106, 2022.

\item L.~Chomiuk and M.S.~Povich, Astron. J. \textbf{142}, 197 (2011).

\item R.F.~Christy, Astrophys. J. \textbf{144}, 108 (1966).

\item P.S.~Conti, Ann. Rev. Astron. Astrophys. {\bf 16}, 371, (1978).

\item J.P.~Cox, A.N.~Cox, K.H.~Olsen, D.S.~King, and D.D.~Eilers, Astrophys. J. \textbf{144}, 1038 (1966).

\item M.R.~Drout, Y.~G\"otberg, B.A.~Ludwig, J.H.~Groh, S.E.~de Mink, A.J.G.~O'Grady, and N.~Smith,
      Science \textbf{382}, 1287 (2023).

\item D.~Dutta and J.~Klencki, Astron. Astrophys. \textbf{687}, A215 (2024).

\item K.~El--Badry and, E.~Quataert, MNRAS \textbf{502}, 343 (2021). 

\item K.~El--Badry and K.~B.~Burdge, MNRAS \textbf{511}, 24 (2022).  

\item K.~El--Badry, C.~Conroy, E.~Quataert,H.--W.~Rix, J.~Labadie--Bartz,
      T.~Jayasinghe, T.~Thompson, P.~Cargile, K.G.~Stassun, and I.~Ilyin, MNRAS \textbf{516}, 3602 (2022). 

\item Yu.A.~Fadeyev, Astron.Lett. \textbf{39}, 306 (2013).

\item Yu.A.~Fadeyev, M.F.~Novikova, Astron. Lett. \textbf{29}, 522 (2003).

\item P.~Giannone and M.A.~Giannuzzi, Astron. Astrophys. \textbf{19}, 298 (1972).

\item Y.~G\"otberg, S.E.~de Mink, J.H.~Groh, T.~Kupfer, P.A.~Crowther, E.~Zapartas, and M.~Renzo,
      Astron. Astrophys. \textbf{615}, A78 (2018).

\item Y.~G\"otberg, M.R.~Drout, A.P.~Ji, J.H.~Groh, B.A.~Ludwig, P.A.~Crowther,
      N.~Smith, A.~de Koter, and S.E.~de Mink,
      Astrophys. J. \textbf{959}, 125 (2023).

\item G.M.H.J.~Habets, Astron. Astrophys. \textbf{165}, 95 (1986).

\item U.~Heber, arXiv:2410.11663 (2024).

\item B.~Hovis--Afflerbach, Y~G\"otberg, A.~Schootemeijer, J.~Klencki, A.L.~Strom,
      B.A.~Ludwig, and M.R.~Drout, arXiv:2412.05356 (2024).

\item J.R.~Hurley, С.A~Tout, and O.R.~Pols MNRAS \textbf{329}, 897 (2002). 

\item I.~Iben~Jr. and A.V.~Tutukov, Astrophys. J. Suppl. Ser. \textbf{58}, 661 (1985).

\item I.~Iben~Jr. and A.V.~Tutukov, Astrophys. J. \textbf{313}, 727 (1987).

\item A.~Irrgang, N.~Przybilla, and G.~Meynet, Nat.Astron., \textbf{6}, 1414 (2022).

\item A.S.~Jermyn, E.B.~Bauer, J.~Schwab, R.~Farmer, W.H.~Ball, E.P.~Bellinger, A. Dotter,
      M.~Joyce, P.~Marchant, J.S.G.~Mombarg, W.M.~Wolf, W.~Sunny, L.~Tin, G.C.~Cinquegrana, E.~Farrell,
      R.~Smolec, A.~Thoul, M.~Cantiello, F.~Herwig, O.~Toloza, L.~Bildsten, R.H.D.~Townsend, and F.X.~Timmes,
      Astrophys. J. Suppl. Ser. \textbf{265}, 15 (2023).

\item R.~Kippenhahn and, A.~Weigert, Zeitschrift f\"ur Astrophys. \textbf{65}, 25 (1967a).

\item R.~Kippenhahn, K.~Kohl., and A.~Weigert, Zeitschrift f\"ur Astrophys. \textbf{66}, 58 (1967b).

\item R.~Klement; T.~Rivinius, D.R.~Gies, D.~Baade, A.~M\'erand, J.D.~Monnier, G.H.~Schaefer,
      C.~Lanthermann, N.~Anugu, S.~Kraus, and T.~Gardner, Astrophys. J. \textbf{962}, 70 (2024).

\item R.~Kuhfu\ss, Astron. Astrophys. \textbf{160}, 116 (1986).

\item  D.~J. Lennon, J.~Ma\'iz Apell\'aniz, A.~Irrgang, R.~Bohlin, S.~Deustua, P.L.~Dufton,
      S.~Sim\'on--D\'iaz, A.~Herrero, J.~Casares, T.~Mu\~noz--Darias, S.J.~Smartt,
      J.I.~Gonz\'alez Hern\'andez, and A.~de Burgos,
      Astron. Astrophys. \textbf{649}, A167 (2021).

\item Y.~Naz{\'e} and G.~Rauw, arXiv:2503.08190  (2025).

\item W.~Packet, Astron. Astrophys. \textbf{102}, 17 (1981).

\item B.~Paczy{\'n}ski, Acta Astron., \textbf{17}, 355 (1967).

\item V.~Ramachandran, J.~Klencki, A.A.C.~Sander, D.~Pauli, T.~Shenar,
      L.M.~Oskinova, and W.--R.~Hamann,
      Astron. Astrophys. \textbf{674}, L12 (2023).

\item V.~Ramachandran, A.A.C.~Sander, D.~Pauli, J.~Klencki, F.~Backs, F.~Tramper,
      M.~Bernini--Peron, P.~Crowther, W.--R.~Hamann, R.~Ignace, R.~Kuiper, M.S.~Oey, L.M.~Oskinova,
      T.~Shenar, H.~Todt, J.S.~Vink, L.~Wang, and A.~Wofford,
      Astron. Astrophys. \textbf{692}, A90 (2024).

\item Т.~Rivinius, R.~Klement, S.D.~Chojnowski, D.~Baade, M.~Abdul--Masih, N.~Przybilla,
      J.~Guarro Fl\'o, B.~Heathcote, P.~Hadrava, P. D.~Gies, K.~Shepard, C.~Buil, O.~Garde, O.~Thizy,
      J.D.~Monnier, N.~Anugu, C.~Lanthermann, G.~Schaefer,C.~Davies, and S.~Kraus,
      Astron. Astrophys. \textbf{694}  A172 (2025). 

\item Y.~Shao and  X.-D~Li, Astrophys. J. \textbf{908}, 67 (2021).

\item T.~Shenar,J.~Bodensteiner, M.~Abdul--Masih, M.~Fabry, L.~Mahy, P.~Marchant,
      G.~Banyard, D.M.~Bowman, K.~Dsilva, C.~Hawcroft, M.~Reggiani, and H.~Sana,
      Astron. Astrophys. \textbf{639}, L6 (2020a). 

\item T.~Shenar, A.~Gilkis, J.S.~Vink, H.~Sana, and A.A.C.~Sander,
      Astron. Astrophys. \textbf{634}, A79 (2020b).

\item J.I.~Villase\~nor, D.J.~Lennon, A.~Picco, T.~Shenar, P.~Marchant, N.~Langer,
      P.L.~Dufton, F.~Nardini, C.J.~Evans, J.~Bodensteiner, S.E.~de Mink, Y.~G\"otberg, Y. I.~Soszy\'nski,
      W.D.~Taylor, and H.~Sana, MNRAS \textbf{525}, 5121 (2023).

\item L.~Wang, D.R.~Gies, G.J.~Peters, and Z.~Han, Astron. J. \textbf{165}, 203 (2023).

\item S.~Woosley, N.~Langer, and Т.А.~Weaver, Astrophys. J. \textbf{448}, 315 (1995).

\item L. Yungelson, A.~Kuranov, K.~Postnov, K. M.~Kuranova, L.M.~Oskinova, and W.--R.~Hamann,
      Astron. Astrophys. \textbf{683}, A37 (2024).

\item J.~Zak, D.~Jones, H.M.J.~Boffin, P.G.~Beck, J.~Klencki, J.~Bodensteiner, T.~Shenar,
      H.~Van Winckel,  K.~Arellano--C\'ordova, J.~Viuho, P.~Sowicka, E.W.~Guenther, and A.~Hatzes,
      MNRAS \textbf{524}, 5749 (2023).

\item J.~Zi{\'o}{\l}kowski, Acta Astron. \textbf{20}, 213 (1970).

\end{enumerate}
}
\end{document}